\documentclass[12pt]{iopart}
\usepackage{iopams}  
\usepackage{epsfig}
\usepackage{graphics}

\def\beq{\begin{equation}}
\def\eeq{\end{equation}}

\def\eps_null{\varepsilon_0}

\begin{document}

\title{Dynamical symmetries and harmonic generation}
\author{F Ceccherini\dag\ , D Bauer\dag\ and F Cornolti\ddag}
\address{\dag\ Theoretical Quantum Electronics (TQE), Darmstadt University of Technology, Hochschulstr.\ 4A, D-64289 Darmstadt, Germany}
\address{\ddag\ Istituto Nazionale di Fisica della Materia (INFM), sez. A, Dipartimento di Fisica, Universit\`a di Pisa, Piazza Torricelli 2, 56100 Pisa, Italy}
\date{\today}

\begin{abstract} 
We discuss harmonic generation in the case of laser field-dressed Hamiltonians that are invariant under so-called dynamical symmetry operations. Examples for such systems are molecules which exhibit a discrete rotational symmetry of order $N$ (e.g. benzene with $N=6$) interacting with a circularly polarized laser field and single atoms in a bichromatic field, with the two lasers having circular polarizations. 
Within  a general group theory approach we study the harmonics one obtains from the interaction of a laser pulse and a circular molecule. When the system is in a pure field-dressed state the known selection rule $kN\pm 1$, $k=1,2,3,\ldots$ results. However, other lines are observed when  recombinations with states of a symmetry different from the initial one become important. This is the case for realistic laser pulses (i.e., with a finite duration). In particular when the fundamental laser frequency (or one of its multiples) is resonant with a transition between field-dressed states.  Numerical {\em ab initio} simulations, confirming our analytical calculations and illustrating the power of the group theory approach, are presented.
\end{abstract}


\submitto{\JPB}
\maketitle

\section{Introduction}
The generation of harmonics by atoms and molecules in a laser field is a topic that has been thoroughly studied during the last fifteen years (see \cite{joachsal} for recent reviews). The big interest lies in the potential of atoms and molecules in laser fields as sources of short-wavelength and short pulse radiation. It is well known that when a strong linearly polarized laser field interacts with an atom only odd harmonics are generated. In more complex systems a detailed study of the symmetry transformations is required for predicting which harmonics are allowed or not \cite{alon,nano}. In a recent work \cite{alon} it was shown that peculiar selection rules can be obtained for molecules which are invariant under a discrete rotational symmetry ${\cal C}_N$ (like benzene where $N=6$). When this kind of molecules interacts with a circularly polarized laser pulse of infinite duration, the selection rules for the harmonics are $n = kN \pm 1$, where $k \in {\cal N}_+$ and $n$ is the harmonic order. This class of molecules is particularly interesting because they can work as a filter, i.e., the first allowed harmonic is $n = N-1$, and therefore, if $N$ is large, the emitted frequency can be very high. All systems whose Hamiltonian is invariant under a ``dynamical rotation'', i.e., a discrete spatial rotation plus a time translation, generate  selection rules of the kind $kN\pm 1$. Together with cyclic molecules another example of such a system is the single atom interacting with two-color laser fields with circular polarizations and frequencies one an integer multiple of the other. In both the molecular and the atomic case it is possible to obtain these selection rules also from the conservation of the angular momentum component perpendicular to the polarization plane. 

In a previous paper \cite{ceccherini} we have presented results from numerical simulations for a benzene model-molecule showing that, when a realistic, finite pulse is taken into account, other lines than those expected from the $kN\pm1$ selection rule can be present. In this paper we will show that, if embedded in a general group theory treatment, also this case can be understood in a systematic way. Our analysis will allow us to achieve a full understanding of the molecule harmonic spectrum. Moreover, we will show that the field-dressed level scheme can be inferred from the harmonic spectrum. The particular features and differences of a finite pulse compared to an infinite one will be also addressed. All the laser fields considered are circularly polarized in the $xy$-plane, and atomic units are used throughout the paper. We apply the dipole approximation which is excellent for the laser frequencies and intensities under consideration.

\section{Dynamical Symmetries}
The Hamiltonian $H_{\rm mo}$ of a cyclic molecule interacting with a single laser field and the Hamiltonian $H_{\rm at}$ of a single atom interacting with two circularly polarized lasers are, from a symmetry point of view, equivalent. Both are invariant under certain dynamical rotations. We will consider  only a single electron active. This assumption does not introduce any loss of generality to our derivations \cite{alon}. Therefore $H_{\rm mo}$ in cylindrical coordinates $(\rho, \varphi, z)$ reads
\begin{eqnarray}
\label{ham_mo}
\nonumber  H_{\rm mo} &=& -\frac{1}{2\rho}\frac{\partial}{\partial \rho}\left( \rho\frac{\partial}{\partial \rho} \right ) - \frac{1}{2\rho^2}\frac{\partial^2}{\partial \varphi^2} - \frac{\partial^2}{\partial z^2} + V_{\rm mo}(\rho, \varphi, z)\\ 
&+& \frac{{\cal E}}{\sqrt{2}}\rho \,\cos (\varphi -\omega t), 
\end{eqnarray} 
where ${\cal E}$ is the amplitude of the electric field. If $V_{\rm mo}(\rho, \varphi, z)$ is a model potential for a cyclic molecule with $N$ ions the Hamiltonian $H_{\rm mo}$ is invariant under the transformation \cite{alon}
\begin{equation}
\label{rotat}
{\hat P}_N = \left ( \varphi \rightarrow \varphi + \frac{2\pi}{N}, t
  \rightarrow t + \frac{2\pi}{N \, \omega} \right ).
\end{equation}

In the case of the single atom in the two-color field $H_{\rm at}$ is
\begin{eqnarray}
\label{ham_at}
\nonumber H_{\rm at} &=& -\frac{1}{2\rho}\frac{\partial}{\partial \rho}\left( \rho\frac{\partial}{\partial \rho} \right ) - \frac{1}{2\rho^2}\frac{\partial^2}{\partial \varphi^2} - \frac{\partial^2}{\partial z^2} + V_{at}(\rho, z) \\
&+& \frac{{\cal E}_1}{\sqrt{2}}\rho \,\cos (\varphi -\omega t)  + \frac{{\cal E}_2}{\sqrt{2}}\rho \,\cos(\varphi + \eta\,\omega t),
\end{eqnarray} 
where ${\cal E}_1$ and ${\cal E}_2$ are the electric field amplitudes of the two laser fields and opposite polarization has been assumed. The second laser has a frequency that is $\eta$ times the frequency of the first laser, with $\eta$ integer. It is nice to see that it is really the second laser which generates the particular symmetry. In fact, all the terms except the last one of $H_{\rm at}$, are invariant under the continuous transformation 
\begin{equation}
{\hat P}_{\infty} = \left ( \varphi \rightarrow \varphi + \theta, 
 t \rightarrow t + \frac{\theta}{\omega} \right ),
\end{equation}
where $\theta$ has, so far, an arbitrary value. If no other term is present in $H_{\rm at}$ no harmonics are generated. Forcing also the last term to be invariant under such a transformation leads to
\begin{equation}
\theta + \eta\,\theta = (\eta+1) \theta = 2\pi k \Longrightarrow \theta = \frac{2 \pi k}{\eta + 1}.
\end{equation}
Therefore the Hamiltonian $H_{\rm at}$ is invariant under the discrete transformation 
\begin{equation}
{\hat P}_{\eta+1} = \left ( \varphi \rightarrow \varphi + \frac{2\pi}{\eta+1}, 
 t \rightarrow t + \frac{2\pi}{(\eta+1)\,\omega} \right ),
\end{equation}
and this is equivalent to the case of a cyclic molecule with $N = \eta+1$. Hence the selection rules for this case are equal to those derived in \cite{alon} for the cyclic molecules, $n = kN \pm 1$. The case where the two lasers have the same polarization can be handled with the same method, leading to similar selection rules. Further details will be discussed elsewhere \cite{ceccherini2}. In the following the two lasers are always considered with opposite polarization.

An experiment with the two frequencies $\omega$ and $2\omega$  was performed \cite{eichmann}, and its theoretical aspects were recently discussed within a semiclassical three-step model \cite{milos00}. Selection rules for the case of an atom interacting with a circularly polarized field of frequency $\omega$ and a linearly polarized field of frequency $N\omega$ were also derived \cite{aver},  confirming numerical results obtained previously \cite{tong}. 

In the derivation of the $kN\pm1$-selection rule one assumes that only a single non degenerate, bound state is involved in the harmonic  generation process. With this restriction the selection rules can be also derived simply with a change of reference frame. In the rotating laser field frame the observer ``sees'' an ionic potential oscillating with a frequency $N\omega$ in a static field; in the two-color atomic case instead, after the change of frame, the atom is at rest in the static field and it experiences an oscillating field of frequency $N \omega$ with $N = \eta + 1$. Hence, in this frame the dipole has only Fourier components of frequency $kN\omega$ with $k \in {\cal N}$. Going back to the laboratory frame where the molecule is at rest, the Fourier components are shifted $\pm \omega$ and the selection rules presented in \cite{alon} are obtained. Keeping the assumption of a single bound state involved in the process, it is, moreover, possible to derive the same selection rules also arguing about the angular momentum only. Let us first discuss about the cyclic molecule: because of the symmetry of the molecule, an expansion of the ground state in spherical harmonics $Y_l^m$ has to look like $\Psi_0 =  \sum_{k=0}^\infty \sum_{l \geq kN} a_{k,l} Y_l^{kN}+ \mbox{c.c.}$, i.e, the ground state is expanded over functions that have the required azimuthal symmetry. The wavefunction is then invariant under a rotation of $2\pi/N$. Let us assume that the laser field has a polarization $\sigma^+$, i.e.,  the projection of the angular momentum of a photon on the field propagation direction is equal to 1. If a harmonic of polarization $\sigma^+$ is emitted, the molecule,  for keeping the symmetry, hence has to absorb $kN+1$ photons so that the final change in the projection of the angular momentum is $kN$. On the contrary, if the harmonic emitted has polarization $\sigma^-$, the molecule must absorb $kN-1$ photons. The consequence is that for each $k$ the higher harmonic $kN+1$ has the same polarization as the laser field, and the lower $kN-1$ the opposite one, as it was obtained with the dynamical symmetry  method in \cite{alon}. For the atom the derivation is similar \cite{becker,milos}. Let us assume an electron is in an $s$ state (the quantum numbers $l$ and $m$ are zero). Let the first laser of frequency $\omega$ have a polarization $\sigma^+$ and the second laser of frequency $\eta\,\omega$ a polarization $\sigma^-$. If a harmonic $\sigma^+$ is emitted the sum of all the components along $z$ of the angular momentum carried by absorbed photons has to be $+1$. If from the first laser the atom absorbs $p$ photons, it must absorb $p-1$ from the second laser. The total absorbed energy is $p\omega + (p-1)\eta \omega = \omega [p(1+\eta) - \eta] = \omega [(p-1)(1+\eta) +1]$. Since $p$ is arbitrary, with $k=p+1$ and $N=\eta +1$ we see that the $(kN+1)^{\rm th}$ harmonic is emitted. With the same argument, starting from a harmonic of polarization $\sigma^-$, we  obtain $kN-1$. 

\section{Group Theory}
From a group theory point of view the Hamiltonian $H_{\rm mo}^0$  of a cyclic molecule without the laser field has a symmetry described by the group ${\cal D}_{Nh}$ \cite{landau,tinca}. When the molecule interacts with a laser field which has no temporal envelope its total Hamiltonian, given by (\ref{ham_mo}),  can be written as $H_{\rm mo} = H_{\rm mo}^0 + W(t)$ where $W(t)\propto \rho \cos (\varphi - \omega t)$ is the molecule-laser interaction. 

First we study the symmetry group of the total Hamiltonian $H_{\rm mo}$. $H_{\rm mo}^0$ has the symmetry ${\cal D}_{Nh}$ and $W$ is described by ${\cal G}_\infty$, i.e, an extension to infinity of the so called ``cyclic dynamical symmetry group'' \cite{alon}, defined by ${\cal G}_L \equiv \{\hat{P}_L, \hat{P}_L^2,\ldots, \hat{P}_L^{L-1},  \hat{P}_L^L = \bf{I} \}$ where $L$ is the order of the group and the number of its elements. $\hat{P}_L$ is the operator already introduced in (\ref{rotat}) which acts in space and time, and is defined as 
\begin{equation}
{\hat P}_L = \left ( \varphi \rightarrow \varphi + \frac{2\pi}{L}, t
  \rightarrow t + \frac{2\pi}{L \, \omega} \right ).
\end{equation}
Combining  ${\cal D}_{Nh}$ and  ${\cal G}_\infty$ we have that the symmetry group of the Hamiltonian $H_{\rm mo}$ is ${\cal G}_{N}$ where $N$ is the number of ions in the molecule which determines the discrete rotational symmetry.

Because ${\cal G}_N$ is a cyclic group, it is also Abelian, and therefore it has the following properties: ({\em i}) each element forms a class by its own and hence there are $N$ irreducible representations, ({\em ii}) each irreducible representation is one-dimensional, i.e., the elements are complex numbers. The group ${\cal G}_N$ is isomorphic to the well known group ${\cal C}_N$ and  therefore the same character table apply. It is interesting to look how, from a physical point of view, the symmetry of the molecular Hamiltonian changes when the laser field is present. The group ${\cal D}_{Nh}$ contains $4N$ elements, i.e., $N$ rotations about an axis of the $N$th order, $N$ rotations through an angle $\pi$ about horizontal axes, $N$ reflections $\sigma_v$ at the vertical planes and $N$ rotary-reflection transformations $C_N^k\sigma_h$ \cite{landau}. Looking at the field-dressed Hamiltonian (\ref{ham_mo}) it appears that only the rotations about the vertical axis are still symmetry transformations. In fact, the reflection at the vertical planes and the rotations about horizontal axes would require a space-dependent time transformation. The reflection $\sigma_h$ at the horizontal plane would change the polarization of the laser field. Hence from the $4N$ possible operations we had for $H_{\rm mo}^0$ only $N$, i.e., the rotations about the vertical axis, keep the Hamiltonian $H_{\rm mo}$ invariant.

For a certain group ${\cal C}_N$ the type of its irreducible representations depends on $N$. If $N$ is odd there are one real and $N-1$ complex representations, if $N$ is even there are two real and $N-2$ complex representations. Half of the complex representations is the complex conjugate of the other half. Usually in the literature, the two complex conjugated representations are considered equivalent and indicated with E, normally reserved for two-dimensional representations \cite{landau}. In our case, because the Hamiltonian (\ref{ham_mo}) is not time-reversal invariant, these representations are not equivalent. Therefore we have to keep all the $N$ representations separated. 

We call each of the $N$ irreducible representations $\mbox{R}_\ell$ where $\ell$ is an index between $0$ and $N-1$. The generator of the $\ell$-th representation is given by $\rme^{\rmi 2\pi \ell/N}$. Therefore, it  follows that the $(m+1)$-th entry of the $\ell$-th representation is $\rme^{\rmi 2\pi \ell m/N}$ (the first entry is given by the identity). Because of the unidimensionality of the representations the product of two representations, $\mbox{R}_\alpha$ and  ${\mbox{R}_\beta}$, is the irreducible representation $\mbox{R}_{\alpha + \beta}$. If $\alpha + \beta \geq N$, $\mbox{R}_{\alpha + \beta}$ is equivalent to $\mbox{R}_{\alpha + \beta - N}$. Thus each entry of the product table is again a representation $\mbox{R}_\ell$ with $0 \leq \ell < N-1$. 

Three conclusions can be drawn now: (i) from the definition of the representation generator we have that the complex conjugated representation of $\mbox{R}_\ell$ is $\mbox{R}_{N-\ell}$, (ii) the representation  $\mbox{R}_0$ has all characters equal to  1 (this is the well known ``totally symmetric irreducible representation'') and it is present for every $N$, (iii) when $N$ is even $\mbox{R}_{N/2}$ is the other real representation. 

The description given so far in terms of group theory  will prove particularly useful later on when we analyze the transitions between the field-dressed states of the molecule. Indeed,  whether an electronic transition between two states, each of them corresponding to an irreducible representation, is allowed or not can be determined examining the product of the symmetries of the initial state, the inducing operator, and the complex conjugated of the final state. If the product leads to the totally symmetric irreducible representation the transition is possible, i.e.,   $\mbox{R}_{\rm fin}^* \bigotimes \mbox{R}_{\rm op} \bigotimes  \mbox{R}_{\rm ini} = \mbox{R}_{0}$ in our notation.

\section{Harmonic generation}
Provided the laser pulse is sufficiently adiabatic and the emitting molecules can be considered uncorrelated it is a good approximation to calculate the harmonic spectrum from the Fourier-transformed dipole expectation value.  With the wavefunction $\Psi(\vec{r}, t)$ we can write the probability to get the $n$th harmonic as 
\begin{eqnarray}
\label{proba}
\nonumber \chi(n) &\propto& n^4 \left | \int \rme^{\rmi n\omega t} \mbox{d}t \int \Psi^* (\vec{r}, t)\rho \rme^{\mp \rmi \varphi} \Psi(\vec{r}, t) \mbox{d}\vec{r} \right |^2\\
&=&  n^4 \left |\int \int \Psi^*(\vec{r}, t) \rho \rme^{\rmi (n \omega t \mp \varphi)} \Psi(\vec{r}, t) \mbox{d}\vec{r} \,\mbox{d}t \right |^2.  
\end{eqnarray}
For a system whose Hamiltonian is periodic in time with a period $\tau = 2\pi/\omega$ we can apply  the Floquet theorem (see, e.g., \cite{faisalbook}). The wavefunction of a certain state of the system can be  written as $\Psi_i(\vec{r},t) = \rme^{-\rmi \xi_i t}\Phi_i(\vec{r},t)$  where $\xi_i$ is the so-called quasi-energy and $\Phi_i(\vec{r}, t+\tau) = \Phi_i(\vec{r}, t)$.  The functions $\Phi_i(\vec{r},t)$ are eigenfunction of ${\cal H}_{\rm mo} \equiv [H_{mo} - \rmi \frac{\partial}{\partial t}]$, i.e., 
\begin{equation}
{\cal H}_{\rm mo}\Phi_i(\vec{r},t) = \xi_i \Phi_i(\vec{r},t). 
\end{equation}
In what follows we refer to $\Phi_i(\vec{r},t)$ as a Floquet state. If the system is in a pure Floquet state we can derive the selection rules as presented in \cite{alon}. However, due to pulse shape effects or other perturbations the system might be not in a pure Floquet state, but rather in a superposition of them, 
\begin{equation}
\Psi(\vec{r},t) = \sum_i \beta_i \rme^{-\rmi \xi_i t} \Phi_i(\vec{r},t). 
\end{equation}
Here, for simplicity, we assume the $\beta_i$s constant (or slowly varying) in time. The quasi-energies are given by $\xi_i = \Delta_i - \rmi \Gamma_i$ where the real part $\Delta_i$ is the field-dressed energy of the state and the imaginary part $\Gamma_i$ corresponds to half the ionization rate. The latter would lead to finite harmonic line widths of Lorentzian shape. In what follows we will neglect this effect keeping only the real part of $\xi_i$ for our symmetry analysis. Eq. (\ref{proba}) then becomes 
\begin{eqnarray}
\label{expan}
\nonumber \chi(n) &\propto& n^4 \Bigg | \sum_i |\beta_i|^2 \int \int  \Phi_i \rho \rme^{\rmi (n \omega t \mp \varphi)}  \Phi_i  \, \mbox{d}\vec{r} \mbox{d}t \\ 
\nonumber &+& \sum_{i\geq j} \beta_i^*\beta_j \int \int  \Phi_i \rme^{\rmi \Delta_i t} \rho \rme^{\rmi (n \omega t \mp \varphi)} \rme^{-\rmi \Delta_j t} \Phi_j \, \mbox{d}\vec{r} \mbox{d}t \\
 &+& \sum_{i\geq j} \beta_j^*\beta_i \int \int  \Phi_j \rme^{\rmi  \Delta_j t} \rho \rme^{\rmi (n \omega t \mp \varphi)} \rme^{-\rmi\Delta_i t} \Phi_i \, \mbox{d}\vec{r} \mbox{d}t  \Bigg|^2. 
\end{eqnarray}
We introduce $\Delta_i^j \equiv \Delta_j - \Delta_i$ and the following operator
\[
\nonumber \hat{G}_\mp^{(i,j)} \equiv \rho \rme^{\rmi ((n \omega -\Delta_i^j) t \mp \varphi)}. 
\]
Noting that the functions $\Phi_i(\vec{r},t)$ form an extended Hilbert space \cite{sambe} Eq.~(\ref{expan}) can be rewritten as 
\begin{eqnarray}
\label{expan2}
\nonumber \chi(n) \propto &n^4& \Bigg | \sum_i |\beta_i|^2 \langle \langle \Phi_i|\hat{G}_\mp^{(i,i)}| \Phi_i \rangle \rangle + \sum_{i\geq j} \beta_i^*\beta_j \langle \langle \Phi_i |\hat{G}_\mp^{(i,j)}| \Phi_j \rangle \rangle \\ 
 &+& \sum_{i\geq j} \beta_j^*\beta_i \langle \langle \Phi_j|\hat{G}_\mp^{(j,i)}| \Phi_i \rangle \rangle \Bigg|^2
\end{eqnarray}
where the double brackets stand for integration over time and space. 

We can think about the different operators $\hat{G}_\mp$ as operators which induce transitions where the initial and the final states can be the same or different. Such transitions are accompanied by the emission of harmonics. If the system is stable in one Floquet state the expansion in (\ref{expan2}) reduces to only one term. In particular, when we consider the  system being in the Floquet ground state $\Phi_0(\vec{r},t)$, $\chi(n)$ is proportional to the Fourier-transform of the expectation value of the dipole calculated with the Floquet ground state, i.e.,  
\begin{equation} 
\chi(n) \propto n^4  \left | \langle \langle \Phi_0|\hat{G}_\mp^{(i,i)}| \Phi_0 \rangle \rangle \right |^2. 
\end{equation}
From (\ref{expan2}),  the structure of the dipole spectrum can be derived. Because $\hat{P}_N^{-1} \hat{P}_N = 1$,
\begin{equation}
\label{cond}
\langle \langle \Phi_i | \hat{G}_\mp^{(i,j)}| \Phi_j \rangle \rangle = 
\langle \langle \hat{P_N} \Phi_i | \hat{P}_N \hat{G}_\mp^{(i,j)} \hat{P}_N^{-1}| \hat{P}_N \Phi_j \rangle \rangle,  
\end{equation}
and writing the operator $\hat{G}_\pm^{(i,j)}$ explictly we have
\begin{eqnarray}
\label{gener}
\nonumber \hat{P}_N \hat{G}_\mp^{(i,j)} \hat{P}_N^{-1} &=& \hat{P}_N \,\rho \, \rme^{\rmi ((n\omega - \Delta_i^j) t \mp \varphi)} \hat{P}_N^{-1} = \rho \, \rme^{\rmi ( (n\omega -\Delta_i^j) t \mp \varphi)} \left [\rme^{\rmi  \frac{2\pi }{N}(n -\frac{\Delta_i^j}{\omega} \mp 1)}\right ] \\
&=& \hat{G}_\mp^{(i,j)} \left [\rme^{\rmi  \frac{2\pi }{N}(n -\frac{\Delta_i^j}{\omega} \mp 1 )}\right ] . 
\end{eqnarray}
Applying $M$ times $\hat{P}_N$ to the operator $\hat{G}_\mp^{(i,j)}$ we obtain 
\begin{equation} 
\label{power}
[\hat{P}_N]^M \,\, \hat{G}_\mp^{(i,j)} [\hat{P}_N^{-1}]^M  = \hat{G}_\mp^{(i,j)} \left [\rme^{\rmi  \frac{2\pi M}{N}(n - \frac{\Delta_i^j}{\omega} \mp 1)} \right ]. \end{equation}
From a group theory point of view the term in square brackets in (\ref{gener}) is simply the generator of the corresponding irreducible representation. From (\ref{power}) we see  how the harmonic generation operator behaves under the dynamical rotations of the symmetry group of the Hamiltonian. Given the irreducible representations of the states $\Phi_i(\vec{r},t)$ and $\Phi_j(\vec{r},t)$ we can derive to which representation the operator $\hat{G}_\mp^{(i,j)}$ must correspond in order to satisfy Eq.(\ref{cond}). In representation notation, we must have $\mbox{R}_{i}^* \bigotimes \mbox{R}_{\rm op} \bigotimes  \mbox{R}_{j} = \mbox{R}_{0}$, where $\mbox{R}_{\rm op}$ is the irreducible representation of the operator $\hat{G}_\mp^{(i,j)}$. From the definition of the character table of an Abelian group \cite{landau} we know that $\hat{P}_N \Phi_j(\vec{r},t)$ is equal to the phase $\rme^{\rmi \frac{2\pi\ell_j}{N}}$ times $\Phi_j(\vec{r},t)$ and similar for $\Phi_i(\vec{r},t)$. Also $\hat{P}_N \hat{G}_\mp^{(i,j)} \hat{P}_N^{-1}$ generates a phase. For allowing the emission of a harmonic the product of these three phases must be the real unity. Consequently, the operator $\hat{G}_\mp^{(i,j)}$ has to belong to the representation $\mbox{R}_{\ell_i - \ell_j}$. Imposing the term in square brackets of Eq. (\ref{gener}) to be equal to the generator of the representation $\mbox{R}_{\ell_i-\ell_j}$ leads to 
\begin{equation}
\rme^{\rmi  \frac{2\pi }{N}(n - \frac{\Delta_i^j}{\omega} \mp 1 )} = \rme^{\rmi \frac{2\pi (\ell_i -\ell_j)}{N}} \Longrightarrow \frac{2\pi }{N}(n - \frac{\Delta_i^j}{\omega} \mp 1 ) = \frac{2\pi (\ell_i-\ell_j)}{N} + 2\pi k.  
\end{equation}
From this follows
\begin{equation}
\hat{G}_\mp^{(i,j)} \quad \longrightarrow \quad n = k\,N  + \ell_j^i + \frac{\Delta_i^j}{\omega} \pm 1
\label{rules}
\end{equation}
where $\ell_j^i = \ell_i - \ell_j$. Introducing the extended index $\tilde{\ell}_i \equiv \ell_i - \frac{\Delta_i}{\omega}$, i.e., the symmetry index minus the field-dressed energy of the state in frequency units, we can rewrite  (\ref{rules}) in the final form\begin{equation}
\hat{G}_\mp^{(i,j)} \quad \longrightarrow \quad n = k\,N  + \tilde{\ell}_j^i \pm 1
\label{rules2}
\end{equation}
where $\tilde{\ell}_j^i = \tilde{\ell}_i - \tilde{\ell}_j$. 

Varying $n$ we force the generic operators $\hat{G}_\mp$  to behave under the dynamical rotations according to the different representations $\mbox{R}_\ell$ of the symmetry group, and for each $\tilde\ell$ certain harmonics are emitted. For the special case $\hat{G}_\mp^{(i,i)}$ one obtains $n = kN \pm 1$, reproducing the results in \cite{alon}. They have been calculated taking $\ell_j^i = 0$ and $\Delta_i^j=0$ because the initial and the final state are the same.  $\hat{G}_\mp^{(i,i)}$ behaves according to the totally symmetric representation $\mbox{R}_0$. 

It is worth to be noted that if the energy difference of the field-dressed states corresponding to the irreducible representations $\mbox{R}_{\ell_i}$  and $\mbox{R}_{\ell_f}$ is equal to $\omega(\ell_i - \ell_f)$ we have a kind of accidental degeneracy: the lines corresponding to such transition cannot be distinguished from those arising from the single state processes. 

The operators $\hat{G}_\mp$  can behave according to any representation. Hence all couplings between the Floquet states of the molecule are in principle possible. This of course does not mean that they have equal probability to occur. Considering that the ground state corresponds to $\mbox{R}_0$, and the other $\mbox{R}_\ell$ represent excited states we can derive the following symmetry properties for the relevant single transitions: ({\em i}) $\mbox{R}_0 \rightarrow \mbox{R}_0$: symmetry  $\mbox{R}_0$,  ({\em ii}) $\mbox{R}_0 \rightarrow \mbox{R}_\ell$: symmetry  $\mbox{R}_\ell$,  ({\em iii}) $\mbox{R}_\ell \rightarrow \mbox{R}_0$: symmetry  $\mbox{R}_\ell^* \equiv \mbox{R}_{N-\ell} \equiv \mbox{R}_{-\ell} $, ({\em iv}) $\mbox{R}_\ell \rightarrow \mbox{R}_\ell$: symmetry  $\mbox{R}_{0}$.

It is evident that when the frequency $\omega$ (or one of its integer multiples) of a finite laser pulse is nearly resonant with a transition to a certain state, this state becomes more easily accessible, and therefore transitions  involving that state are more likely to happen. On the other hand, when the laser frequency is tuned off from all the resonances mostly the field-dressed ground state, i.e. $\mbox{R}_0$, is involved and it will be the only one giving a significant contribution to   harmonic generation.   

The fact that selection rules for different initial and final states contain the terms $\Delta_i$  is particularly interesting for the case where the rotating field removes a degeneracy. The measurement of the distance between the satellite lines arising from the two states gives an estimation of the field-dressed level-splitting.

\section{Benzene Molecule} 
The benzene molecule has a symmetry ${\cal D}_{6h}$, but as long as we consider only functions which are all even or odd with respect to reflection at the molecule plane ($\sigma_h$ in the standard notation) it is sufficient to consider only the symmetry operations of the reduced group ${\cal D}_{6}$. The character table of the group ${\cal D}_6$ is given in Tab.~1.

Connecting the four lowest states of the benzene molecule  and the irreducible representations of the group ${\cal D}_6$, i.e., $\mbox{A}_1$, $\mbox{B}_1$, $\mbox{E}_2$, $\mbox{E}_1$ leads to the following energetic sequence: $\mbox{A}_1$ (ground state), $\mbox{E}_1$ (first excited, two-dimensional, i.e.,  two-fold degenerated), $\mbox{E}_2$ (second excited, two-fold degenerated), $\mbox{B}_1$ (third excited, non-degenerated)  \cite{tinca}. In the ground state $\mbox{A}_1$ and  $\mbox{E}_1$ contain two and four $\pi$-electrons, respectively,  and the other levels are empty \cite{lomont}. As it was discussed in the previous section, when the circularly polarized laser field has an infinite duration, the total Hamiltonian $H_{\rm mo} = H_{\rm mo}^0 + W$ would belong to the dynamical symmetry group ${\cal G}_6$ which is isomorphic to ${\cal C}_6$. The character table of the groups ${\cal C}_6$ and  ${\cal G}_6$ is shown in Tab 2. For the group ${\cal C}_6$ the standard notation is reported, for the group  ${\cal G}_6$ instead we use the notation that has been previously introduced. With the laser field  switched on the degeneracies connected to the representations $\mbox{E}_1$ and $\mbox{E}_2$ are removed. If we take for example the degenerated level ${\mbox E_1}$ of the field-free system, in an infinite pulse it is reducible, and knowing the irreducible representations of the group ${\cal C}_6$ we can easily decompose it into irreducible parts: ${\mbox{E}_1} \longrightarrow {\mbox{E}_1^a} + {\mbox{E}_1^b} = {\mbox{R}_1} + {\mbox{R}_5}$. The same argument can be applied to the level ${\mbox{E}_2}$,  ${\mbox{E}_2} \longrightarrow {\mbox{E}_2^a} + {\mbox{E}_2^b} = {\mbox{R}_2} + {\mbox{R}_4}$.

What is worth to be stressed is that, contrary to the standard case in quantum mechanics textbooks, here the two representations which are complex conjugated, like ${\mbox{R}_1}$, ${\mbox{R}_5}$ and ${\mbox{R}_2}$, ${\mbox{R}_4}$, correspond to two different energy levels and therefore the so-called "physically irreducible'' representations really generate  two independent states. This particular feature is due to the non-invariance of the total Hamiltonian $H$ under a time inversion. A similar case is obtained when a magnetic field is considered.

At this point, with an oscillating field infinite in time, we can think about the field-dressed molecule states in pure terms of ${\mbox R}_i$, where $i$ runs from $0$ to $5$.  This system is stable, i.e., there are no transitions between the states corresponding to different representations, unless external perturbations are present. The selection rules in this case can be derived from the product table, Tab. 3. When the pulse has a finite length  we can consider the pulse shape as a perturbation of a system that has a ${\cal D}_6$ symmetry. A transition between two states is possible only if the operator inducing the transition has the ``proper'' symmetry. As it can be derived from Tab. 1 the representations $\mbox{A}_1$, $\mbox{E}_1$, $\mbox{E}_2$ and $\mbox{B}_1$ correspond to rotations of angle $0$, $\frac{\pi}{3}$, $\frac{2\pi}{3}$ and $\pi$, respectively. In terms of $\ell$ they correspond to $\ell = 0$, $\ell = 1$, $\ell = 2$ and $\ell = 3$.

\section{Numerical Simulations}
In this Section we illustrate and test our group theoretical method by applying it to harmonic spectra obtained with the help of {\em ab initio} numerical simulations.    In order to keep the numerical effort feasible we  restrict ourselves to a two-dimensional (2D) model where the molecule plane and the rotating electric field are properly represented. We study a single active electron in a ring-shaped potential with ${\cal C}_N$ symmetry ($N = 6$ in the benzene case). Results from a fully 3D calculation could be quantitatively different but the structure of the spectrum would be the same. We use the potential \cite{ceccherini}
\begin{equation}   
V(\rho, \varphi) = -\frac{A}{\sqrt{(\rho-r_0)^2 + \beta}}\left [\alpha\,\mbox{cos}\,(N\,\varphi)+2-\alpha \right ]
\end{equation}
where $r_0$ is the radius of the molecule ($r_0 = 2.6$), and $\rho$ and $\varphi$ are  polar coordinates. $\beta$ is the parameter which gives the ``degree of smoothness'' of the potential and determines the width of the ground state along the ring. $\alpha$ moves the local maxima of the potential keeping the minima constant (the parameter $\alpha$ is introduced in order to avoid the presence of points where $V=0$ for finite $\rho$, because that could generate non-physical ionization). Finally, $A$ is the  ``strength'' of the potential. For our simulations we chose $\alpha = 0.99$ and $\beta = 0.38$. We used $A$ for tuning the ionization energy of the molecule. For more details about the model and its properties see \cite{ceccherini}. 

When the correct ionization energy for the benzene molecule is taken, the four states are relatively weakly bound and hence, as it has been already discussed in \cite{ceccherini}, for inducing a multiphoton process with a low ionization rate a low frequency is required. As the main aim of the numerical simulations presented in this work is to test {\em ab initio} the analytical derivations discussed in the previous sections, we chose $A = 1.0$ which is less demanding for what concerns the run-time because a higher frequency can be used \cite{ceccherini}. 

If with $\tilde{\mbox{R}}$ we indicate the field-free energy of the level corresponding to the ${\mbox{R}}$ representation of the ${\cal D}_6$ group, with $A = 1.0$ the energies of the four states are: $\tilde{\mbox{A}}_1 = -1.292$, $\tilde{\mbox{E}}_1 = -1.226$, $\tilde{\mbox{E}}_2 = -1.098$ and  $\tilde{\mbox{B}}_1 = -1.027$, and the energy gaps between the excited states and the ground state are $\Omega_1 = 0.0658$, $\Omega_2 = 0.1937$ and $\Omega_3 = 0.2648$. In order to have a convenient way of referencing we introduce the notation ${\bf L}_{\pm}^{ij}$ for indicating a transition between the state ${\mbox{R}}_i$ and ${\mbox{R}}_j$ and its corresponding line in the spectrum. ${\bf L}$ is the $\ell$ which gives the transition symmetry. The lower index distinguishes between the two signs in equation (\ref{rules}). The upper indices specify the initial and the final state, respectively. The order of the states in the upper index gives information about the sign of the line shift due to the distance between the states. Most of the non-single-state transitions discussed here connect  an excited state with the ground state. Transitions between excited states are possible but much less likely to occur. The position of the lines due to a ``transition'' from the ground to an excited state is ${\bf L}_{\pm}^{0i} = kN + i -\Delta_i \pm 1 $, and for the opposite process ${\bf L}_{\pm}^{i0} = kN - i +\Delta_i \pm 1 $, where $\Delta_i$ is the  distance between the ground state and the state ${\mbox{R}}_i$. 

When the laser is switched on the states move because of the dynamical Stark effect and therefore $\Delta_i$ is given by $\Delta_i = \Omega_i + \delta_i$, where $\Omega_i$ is the separation between the two states when no field is present and $\delta_i$ is the relative shift due to the presence of the oscillating field.   

At this point, as we know the selection rules and the values of the field-free energy intervals $\Omega_i$, we can predict the structure of a spectrum where also non-single state processes are involved. The only uncertainty is due to the values of the field shifts $\delta_i$. However, those shifts can be determined by performing several simulations with different laser intensities: when the field is low $\delta_i$ is small (in the limit of zero field $\Delta_i \longrightarrow \Omega_i$) and the lines are located in the proximity of the position expected from the unperturbed level scheme. With increasing field the shift of the states can be followed. Those shifts can become relatively large but the interpretation of the spectrum remains always unambiguous. 

An example of a spectrum, where  also lines occurring from processes between different states are present, is shown in Fig.~1. There, a sine-square pulse of 40 cycles and frequency $\omega = 0.0942$ interacting with the model molecule has been simulated. The peak field strength $\hat{\cal E}$ was 0.14. It is possible to figure out  for each peak the transition which generates it. The same symbol indicates the same process but with different multiplier $k$, i.e., different number of photons. A certain process is repeated every $N = 6$ frequency units. The states that together with the ground state play a role are the first and the second excited, i.e., ${\mbox{E}}_1$, and ${\mbox{E}}_2$. The first state generates transitions of type ${\bf 1}$, the second state of type ${\bf 2}$. With the parameters used for the simulation of Fig. 1 the role of the second state is more important than that one of the first excited state. This is mainly related to how strong the resonance with a particular state is. With a different frequency the relative intensity of the lines connected to the two processes can differ strongly. From the fact that the degeneracies of the states ${\mbox{E}}_1$ and ${\mbox{E}}_2$ are removed one could expect more lines than present in Fig.~1.  Let us discuss for example about the states ${\mbox{E}}_2$, that in our case are the most important  role players. Each of the states ${\mbox{E}}_2$ for a given $k$ generates two lines given by ${\bf 2}_{\pm}^{02} = kN + 2 -\Delta_2\pm 1$ and two given by ${\bf 2}_{\pm}^{20} = kN - 2 +\Delta_2\pm 1$. Therefore eight lines are expected. However, using $\mbox{E}_2^a$ and $\mbox{E}_2^b$ to distinguish the two states $\mbox{E}_2$,  we find the  shifts $\Delta_2^a = \Omega_2^a +\delta_2^a$ and $\Delta_2^b = \Omega_2^b +\delta_2^b$. Considering that $\Omega_2^a = \Omega_2^b$ it follows that if the difference of the shifts $\delta_i$ due to the laser field is very small the eight lines are gathered in four couples, each of them containing two very close lines, difficult to resolve. Moreover, what makes the distinction between the two lines in a couple possible  is not the absolute value of $\delta_i$ but the difference between the two shifts. As the two states without field are degenerate, this difference is the splitting  generated by the field. With our notation we have for the first couple of lines, e.g., ${\bf 2}_{-}^{02(b)} - {\bf 2}_{-}^{02(a)} = \delta_2^a - \delta_2^b$. The same applies for the other three couples. This explains why in Fig. 1 only four lines due to the states $\mbox{E}_2$ are observed. 

In order to obtain a larger splitting, more easily detectable, we have to study how the shift $\delta_i$ behaves with respect to the different field parameters. It was found that with lower frequencies the separation among the states which without field would be degenerate becomes more significant, as one would expect from a ponderomotive scaling. Moreover, it has been observed that the value of each $\delta_i$ can become relatively large compared to $\Omega_i$. In Fig. 2 we show how the location of certain lines moves when the field is increased.

The dipole spectrum obtained with the lower frequency $\omega = 0.0785$ (all other parameters the same as before) is shown in Fig. 3.  Each of the lines  in Fig. 1 is splitted in two, confirming our predictions. 

In order to investigate the role of pulse shape effects simulations using trapezoidal pulses with different rampings have been performed. In Fig.~4 two spectra for pulses of the same length but with a $11$ cycle and $3$ cycle ramping, respectively, are reported. It appears that in both case the overall structure is like in the case of the sine-square pulse. It is worth to note that when the ramping is shorter, Fig. 4b, the relevance of the ``extra'' lines with respect to the ${\bf O}$-lines originating from the single state transitions is higher than in the case of longer ramping, Fig 4a. This is due to the fact that when the ramping is shorter the pulse is less adiabatic and therefore transitions between different Floquet states are induced more likely. In the limit of an infinitely long ramping only the lines ${\bf 0}$ would be present.    
 
\section{Conclusions}
We presented a general group theoretical approach to harmonic generation by systems possessing a discrete rotational symmetry ${{\cal C}}_N$, in particular a ring-shaped molecule in a circularly polarized laser field (e.g., benzene).  In the simple case where the initial and the final field-dressed electronic states are the same the known selection rule $kN\pm1$, $k=1,2,3,\ldots$ was obtained. We demonstrated that this selection rule follows also from a change of reference frame as well as angular momentum conservation in a straight forward manner. Since due to this selection rule less harmonics are allowed within a fixed frequency range such systems might be beneficial for generating short wavelength radiation more efficiently, i.e., without wasting a vast amount of laser energy in undesired laser harmonics.  However, we showed that in general other harmonics different from the expected ones at the positions $(kN\pm1)\omega$ are present as well. Those lines carry important information about the laser field-dressed level scheme of the molecule since energy differences between different states are involved. Transitions between field-dressed states occur because the laser pulse has a temporal shape instead of being infinite as assumed in the derivation of the $kN\pm1$-selection rule. For the case of the ring-shaped molecule we showed that our group theoretical method is capable to predict also the position of those extra  lines. This was demonstrated with the help of harmonic spectra obtained by an {\em ab initio} numerical simulation of a benzene model molecule interacting with a finite laser pulse. The extra harmonics are particularly pronounced in the cases where the laser frequency (or a multiple of it) becomes resonant with electronic transitions between field-dressed states. Harmonic peak splittings and shifts were fully understood in the framework of the group theory approach.
From the symmetry point of view the case of a cyclic molecule in a circularly polarized laser field is equivalent to a single atom in a two-color field with one frequency an integer multiple of the other. Of course, the groups representing the unperturbed systems are different.

\ack{This work was supported in part by the Deutsche Forschungsgemeinschaft within the SPP ``Wechselwirkung intensiver Laserfelder mit Materie'' and in part by INFM through the Parallel Computing Initiative (FUMOFIL project) and the Advanced Research Project CLUSTERS.}

\pagebreak

\begin{table}
\caption{Character table of the group ${\cal D}_6$.}
\begin{indented}
\lineup
\item[]\begin{tabular}{@{}r|rrrrrr}
\br
${\cal D}_6$ \quad & \quad $E$ \quad & $C_2$ \quad & $2C_3$ \quad  & $2C_6$ \quad & $3C_2'$ \quad & $3C_2''$ \cr
\mr 
$\mbox{A}_1$ \quad & \quad $1$ &\quad $1$ &\quad $1$ &\quad  $1$ &\quad $1$ &\quad $1$ \cr
$\mbox{A}_2$ \quad & \quad $1$ &\quad $1$ &\quad $1$ &\quad  $1$ &\quad $-1$ &\quad $-1$ \cr
$\mbox{B}_1$ \quad & \quad $1$ &\quad $-1$ &\quad $1$ &\quad  $-1$ &\quad $1$ &\quad $-1$ \cr
$\mbox{B}_2$ \quad & \quad $1$ &\quad $-1$ &\quad $1$ &\quad  $-1$ &\quad $-1$ &\quad $1$ \cr
$\mbox{E}_2$ \quad & \quad $2$ &\quad $2$ &\quad $-1$ &\quad  $-1$ &\quad $0$ &\quad $0$ \cr  
$\mbox{E}_1$ \quad & \quad $2$ &\quad $-2$ &\quad $-1$ &\quad  $1$ &\quad $0$ &\quad $0$ \cr 
\br 
\end{tabular}
\end{indented}
\end{table}

\begin{table}
\caption{Character table of the group ${\cal G}_6$ and its isomorphic group ${\cal C}_6$. }
\begin{indented}
\lineup
\item[]\begin{tabular}{@{}rr|rrrrrr}
\br
${\cal G}_6$ & \quad ${\cal C}_6$ \quad & \quad $E$ \quad & $C_6$ \quad & $C_3$ \quad  & $C_2$ \quad & $C_3^2$ \quad & $C_6^5$ \cr
\mr
$\mbox{R}_0$ &\quad  A \quad & \quad $1$ &\quad $1$ &\quad $1$ &\quad  $1$ &\quad $1$ &\quad $1$ \\
$\mbox{R}_3$ &\quad  B \quad & \quad $1$ &\quad $-1$ &\quad $1$ &\quad  $-1$ &\quad $1$ &\quad $-1$ \\
$\mbox{R}_1$ &\quad  $\mbox{E}_1^a$ \quad & \quad $1$ &\quad $\omega$ &\quad $\omega^2$ &\quad  $\omega^3$ &\quad $\omega^4$ &\quad $\omega^5$ \\
$\mbox{R}_5$ &\quad  $\mbox{E}_1^b$ \quad & \quad $1$ &\quad $\omega^5$ &\quad $\omega^4$ &\quad  $\omega^3$ &\quad $\omega^2$ &\quad $\omega$ \\
$\mbox{R}_2$ &\quad  $\mbox{E}_2^a$ \quad & \quad $1$ &\quad $\omega^2$ &\quad $\omega^4$ &\quad  $1$ &\quad $\omega^2$ &\quad $\omega^4$ \\  
$\mbox{R}_4$ &\quad  $\mbox{E}_2^b$ \quad & \quad $1$ &\quad $\omega^4$ &\quad $\omega^2$ &\quad  $1$ &\quad $\omega^4$ &\quad $\omega^2$ \\  
\br 
\end{tabular}
\end{indented}
\end{table}

\begin{table}
\caption{ Product table of the irreducible representations of the group ${\cal C}_6$ when the degeneracy due to the time inversion is removed.}
\begin{indented}
\lineup
\item[]\begin{tabular}{@{}c|rrrrrr}
\br
$\bigotimes$ \quad & \quad $\mbox{R}_0$ \quad & $\mbox{R}_1$ \quad & $\mbox{R}_5$ \quad  & $\mbox{R}_2$ \quad & $\mbox{R}_4$ \quad & $\mbox{R}_3$ \cr
\mr 
$\mbox{R}_0$ \quad & \quad $\mbox{R}_0$ \quad & $\mbox{R}_1$ \quad & $\mbox{R}_5$ \quad  & $\mbox{R}_2$ \quad & $\mbox{R}_4$ \quad & $\mbox{R}_3$ \\
$\mbox{R}_1$ \quad & \quad $\mbox{R}_1$ \quad & $\mbox{R}_2$ \quad & $\mbox{R}_0$ \quad  & $\mbox{R}_3$ \quad & $\mbox{R}_5$ \quad & $\mbox{R}_4$ \\
$\mbox{R}_5$ \quad & \quad $\mbox{R}_5$ \quad & $\mbox{R}_0$ \quad & $\mbox{R}_4$ \quad  & $\mbox{R}_1$ \quad & $\mbox{R}_3$ \quad & $\mbox{R}_2$ \\
$\mbox{R}_2$ \quad & \quad $\mbox{R}_2$ \quad & $\mbox{R}_3$ \quad & $\mbox{R}_1$ \quad  & $\mbox{R}_4$ \quad & $\mbox{R}_0$ \quad & $\mbox{R}_5$ \\
$\mbox{R}_4$ \quad & \quad $\mbox{R}_4$ \quad & $\mbox{R}_5$ \quad & $\mbox{R}_3$ \quad  & $\mbox{R}_0$ \quad & $\mbox{R}_2$ \quad & $\mbox{R}_1$ \\
$\mbox{R}_3$ \quad & \quad $\mbox{R}_3$ \quad & $\mbox{R}_4$ \quad & $\mbox{R}_2$ \quad  & $\mbox{R}_5$ \quad & $\mbox{R}_2$ \quad & $\mbox{R}_0$ \\
\br 
\end{tabular}
\end{indented}
\end{table}


\begin{figure}
\begin{center}
\epsfbox{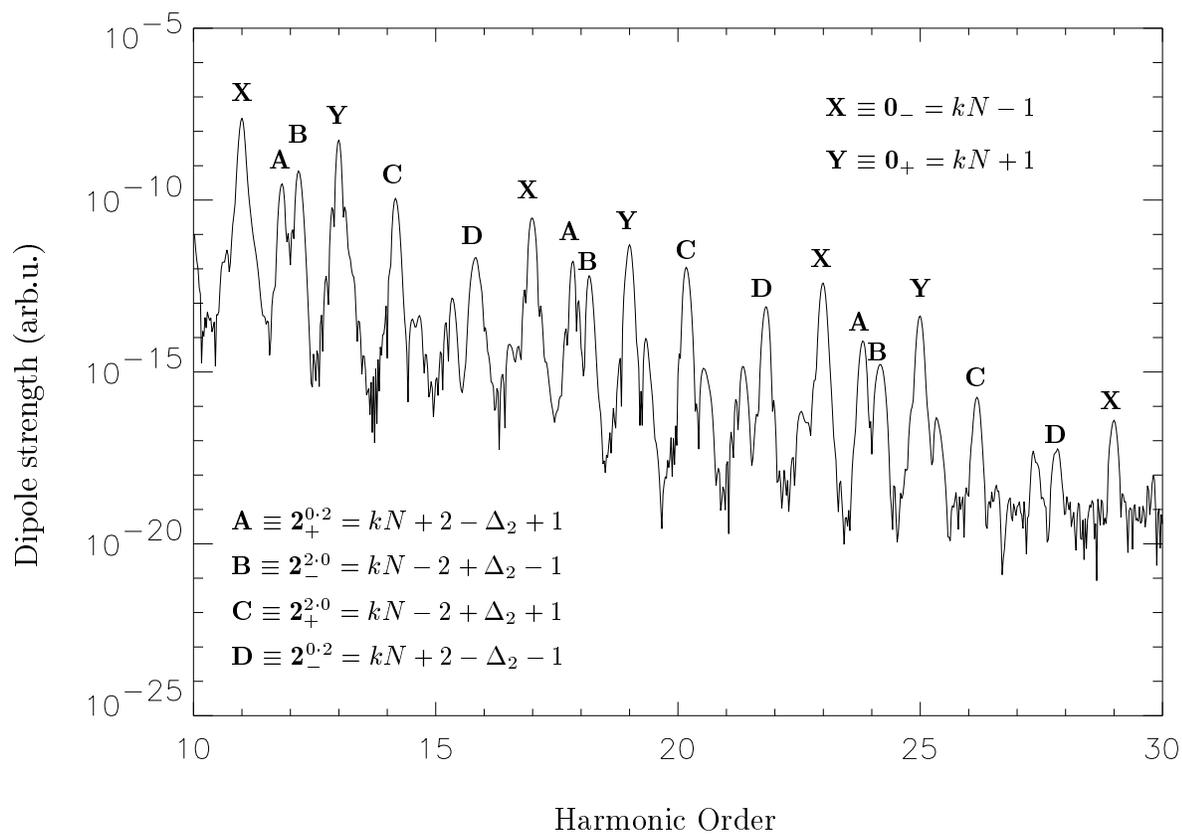}
\end{center}
\caption{A part of the dipole spectrum for a benzene model molecule. The laser frequency was $0.0942$. Together with the lines expected by the single state approach ({\bf X} and {\bf Y}) other lines due to recombinations with states of different symmetries are present. The second excited state plays an important role.}
\end{figure}


\begin{figure}
\begin{center}
\epsfbox{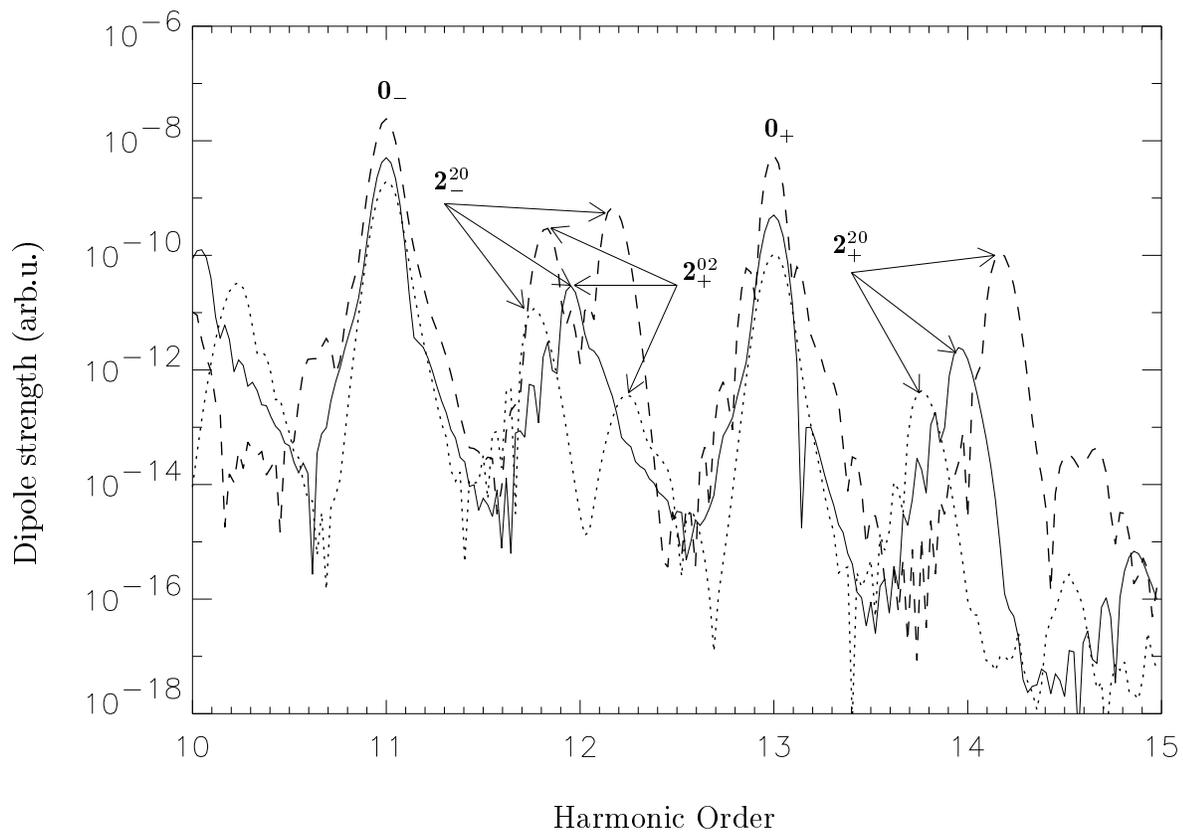}
\end{center}
\caption{Dipole spectra for different laser peak intensities: $\hat{\cal E}=0.10$ (solid line), $\hat{\cal E}=0.08$ (dotted), and $\hat{\cal E}=0.14$ (dashed).  The laser frequency was $0.0942$. The shift of the lines due to the ac Stark effect can be relatively strong. The lines move right or left according to which was the initial state.}
\end{figure}


\begin{figure}
\begin{center}
\epsfbox{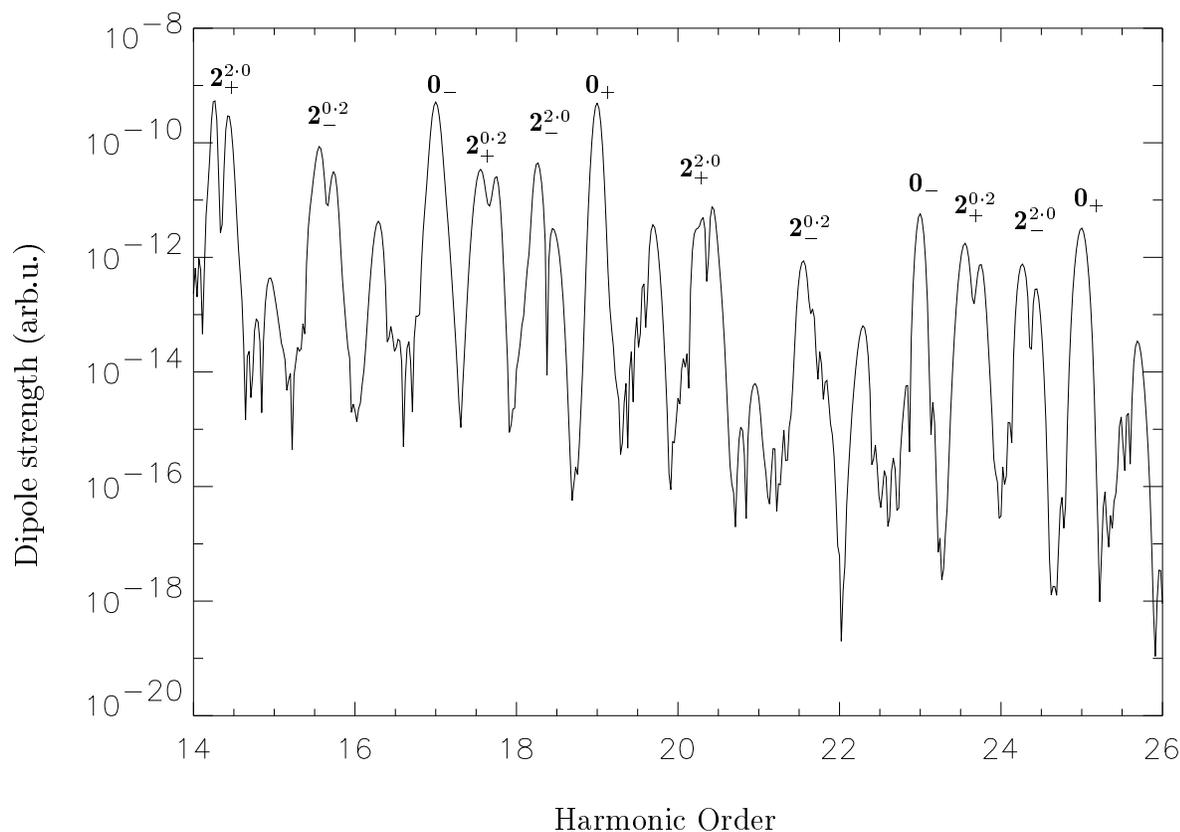}
\end{center}
\caption{Dipole spectrum of the benzene model molecule at a lower laser frequency $\omega=0.0785$ where the peak splitting due to the removal of degeneracies is more clearly visible.}
\end{figure}


\begin{figure}
\begin{center}
\epsfbox{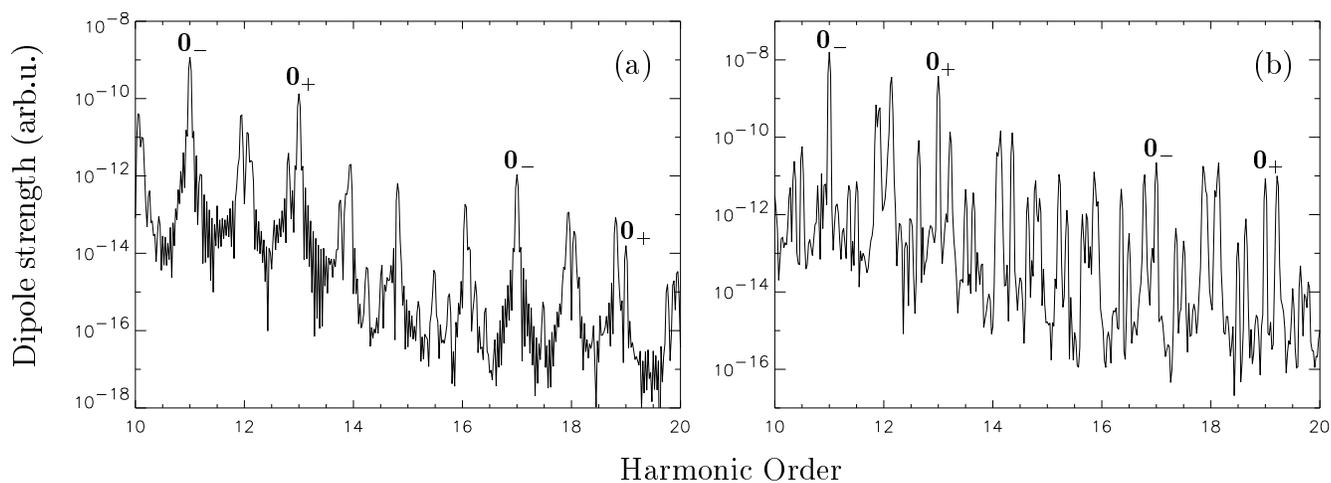}
\end{center}
\caption{A part of the dipole spectra for two trapezoidal pulses of equal total length, $42$ cycles,  but different ramping. In (a) the ramping was $11$ cycles and in (b) $3$ cycles. As expected, when the ramping is shorter, the lines due to transitions between different states are enhanced with respect to those involving a single state, ${\bf 0}_-$ and ${\bf 0}_+$.} 
\end{figure}

\end{document}